\documentclass[aps,prl,twocolumn,superscriptaddress,longbibliography]{revtex4-2}
\usepackage{graphicx}
\usepackage{amssymb,amsmath}
\usepackage{bm}
\usepackage{dcolumn}
\usepackage{float}
\usepackage[OT1]{fontenc} 
\usepackage{url}
\usepackage{mathrsfs}
\usepackage{slashed,comment}
\usepackage{color}
\usepackage{verbatim}
\usepackage{enumitem}
\usepackage{soul,physics}
\usepackage[driverfallback=dvipdfm]{hyperref}
\hypersetup{pdfpagemode=FullScreen,colorlinks=true,breaklinks,urlcolor=blue,linkcolor=blue,citecolor=blue}

\usepackage{subfigure}
\usepackage{amssymb}
\usepackage{bm}
\usepackage{graphicx}
\usepackage{amsmath,amssymb}

\usepackage{pdfpages} 
\usepackage{pgffor} 

\makeatletter
\AtBeginDocument{\let\LS@rot\@undefined}
\makeatother

\def\supplementfilename{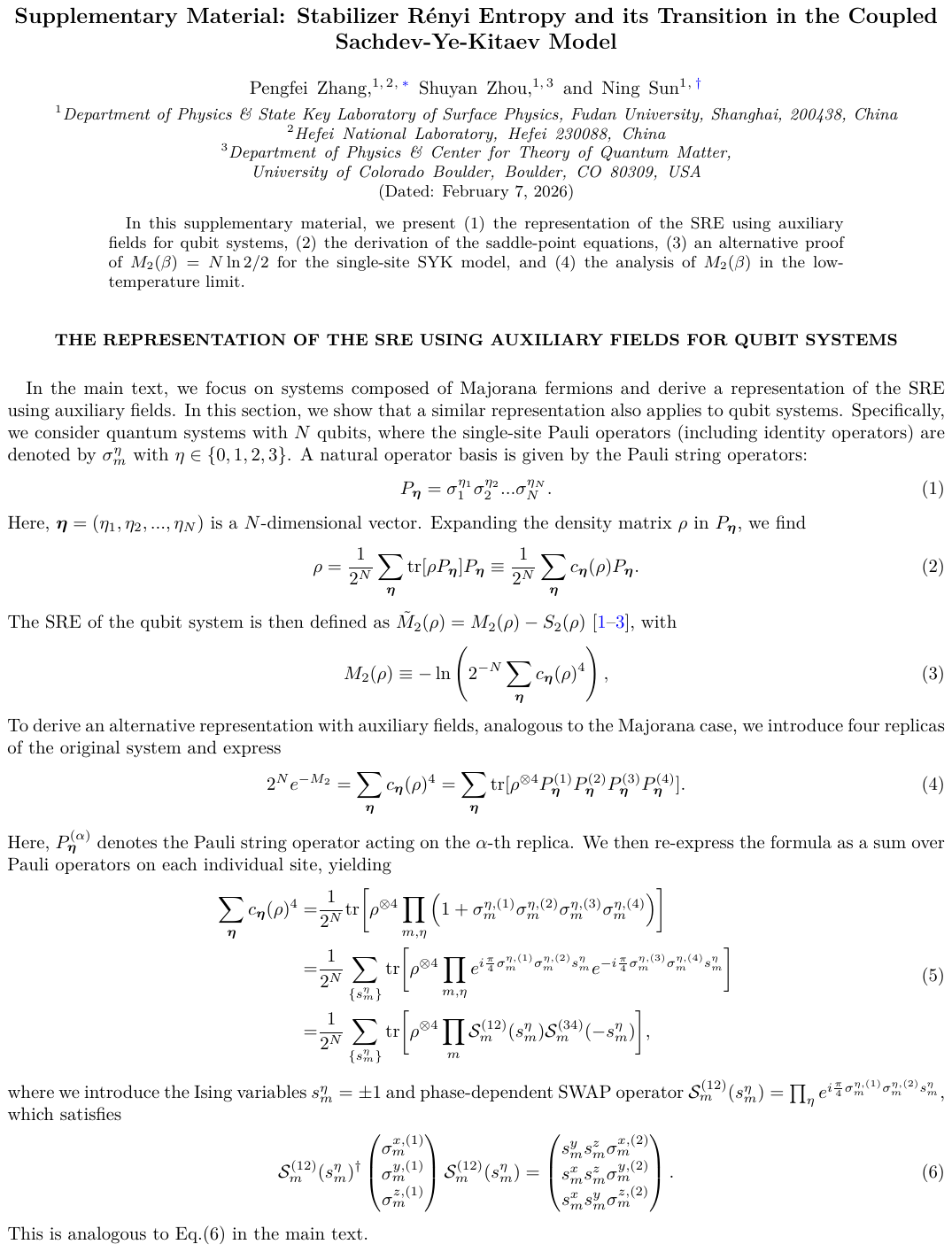}

\pdfximage{\supplementfilename}
\def\numbersupplementpages{\the\pdflastximagepages}

\newif\ifarXiv
\arXivtrue 

\begin{document}
 
  \title{Stabilizer R\'enyi Entropy and its Transition in the Coupled Sachdev-Ye-Kitaev Model}

  \author{Pengfei Zhang}
  \thanks{PengfeiZhang.physics@gmail.com}
  \affiliation{Department of Physics \& State Key Laboratory of Surface Physics, Fudan University, Shanghai, 200438, China}
  \affiliation{Hefei National Laboratory, Hefei 230088, China}

  \author{Shuyan Zhou}
  \affiliation{Department of Physics \& State Key Laboratory of Surface Physics, Fudan University, Shanghai, 200438, China}
  \affiliation{Department of Physics \& Center for Theory of Quantum Matter,
University of Colorado Boulder, Boulder, CO 80309, USA}

  \author{Ning Sun}
  \thanks{ningsun.atom@gmail.com}
  \affiliation{Department of Physics \& State Key Laboratory of Surface Physics, Fudan University, Shanghai, 200438, China}

  \date{\today}

  \begin{abstract}
  Quantum entanglement and quantum magic are two distinct fundamental resources that enable quantum systems to exhibit complex phenomena beyond the capabilities of classical computer simulations. While quantum entanglement has been extensively used to characterize both equilibrium and dynamical phases, the study of quantum magic, typically quantified by the stabilizer R\'enyi entropy (SRE), remains largely limited to numerical simulations of moderate system sizes. In this Letter, we establish a general framework for analyzing the SRE in solvable Sachdev-Ye-Kitaev (SYK) models in the large-$N$ limit, which enables the application of the saddle-point approximation. Applying this method to the Maldacena-Qi coupled SYK model, we identify a series of first-order transitions of the SRE as the temperature is tuned. In particular, we uncover an intrinsic transition of the SRE that cannot be detected through thermodynamic quantities. We also discuss the theoretical understanding of the SRE in both the high-temperature and low-temperature limits. Our results pave the way for studying the SRE in strongly correlated fermionic systems in the thermodynamic limit and suggest a new class of transitions for which the SRE serves as an order parameter.
  \end{abstract}
   
  \maketitle

  \emph{ \color{blue}Introduction.--} Quantum computation provides a promising framework for addressing computational tasks that are intractable for classical computers~\cite{nielsen2010quantum}. Its power fundamentally relies on two distinct resources of quantum systems: quantum entanglement~\cite{RevModPhys.91.025001} and quantum magic~\cite{PhysRevA.71.022316,PhysRevA.86.052329,Emerson:2013zse,Veitch:2012ttw}. These two resources play complementary roles. Quantum states with low entanglement can be efficiently approximated using tensor network methods~\cite{White:1992zz,Orus:2018dya}, yet they may still exhibit extensive quantum magic. Conversely, the Gottesman-Knill theorem states that non-magic stabilizer states can be efficiently represented within the stabilizer formalism~\cite{Gottesman:1997zz,Gottesman:1998hu,PhysRevA.57.127,PhysRevA.70.052328}, even though they are capable of hosting extensive entanglement. This complementarity highlights that entanglement and magic should be regarded as independent, yet intertwined, resources for quantum computational advantage.

  Meanwhile, intuitions from quantum information science also provide deep insights into quantum many-body physics. In particular, quantum entanglement has emerged as a powerful tool: it enables the classification of topological phases of matter~\cite{PhysRevLett.96.110404,PhysRevLett.96.110405}, distinguishes thermalized from localized states~\cite{RevModPhys.91.021001,2018CRPhy..19..498A,2015ARCMP...6..383A,2015ARCMP...6...15N,PhysRevLett.111.127201,PhysRevB.90.174202,PhysRevLett.110.067204}, and even serves as an order parameter for measurement-induced phase transitions~\cite{Li:2018mcv,Skinner:2018tjl,Chan:2018upn,Gullans:2019zdf,Choi:2019nhg}. In contrast, the characterization of quantum magic in many-body states remains relatively unexplored, despite the proposal of various quantitative measures~\cite{2017PhRvL.118i0501H,Liu:2020yso,Beverland:2019jej,PhysRevA.109.L040401,Turkeshi:2023ctq,msm2-vmg7,Ahmadi:2022bkg,Leone:2021rzd,Wang:2023uog,Haug:2023hcs,Qian:2025oit}, including the stabilizer R\'enyi entropy (SRE)~\cite{Leone:2021rzd,Haug:2023hcs,Wang:2023uog}. Recent progress has primarily focused on numerical simulations in interacting quantum systems~\cite{White:2020zoz,Smith:2024ydp,Bera:2025pfp,Jasser:2025myz,njgn-fksh,Odavic:2024lqq,PhysRevA.110.022436,Tarabunga:2023xmv,Tarabunga:2023hau,Catalano:2024bdh,Viscardi:2025vya,Ding:2025nua,PRXQuantum.4.040317} and free-fermion systems~\cite{Oliviero:2022euv,Collura:2024ida,Wang:2025csz} of moderate system sizes, or analyses of matrix product states~\cite{Lami:2023naw,PhysRevB.107.035148,PhysRevLett.133.010602,Tarabunga:2025wym} and random circuits~\cite{Turkeshi:2023lqu,Zhang:2024fyp,Turkeshi:2024pnj,Tirrito:2024kts,Haug:2024ptu,Szombathy:2024tow,Szombathy:2025euv,Hou:2025bau}. A major challenge remains the lack of a general path-integral formulation for studying the SRE, as exists for equilibrium properties.

  \begin{figure}[t]
    \centering
    \includegraphics[width=0.72\linewidth]{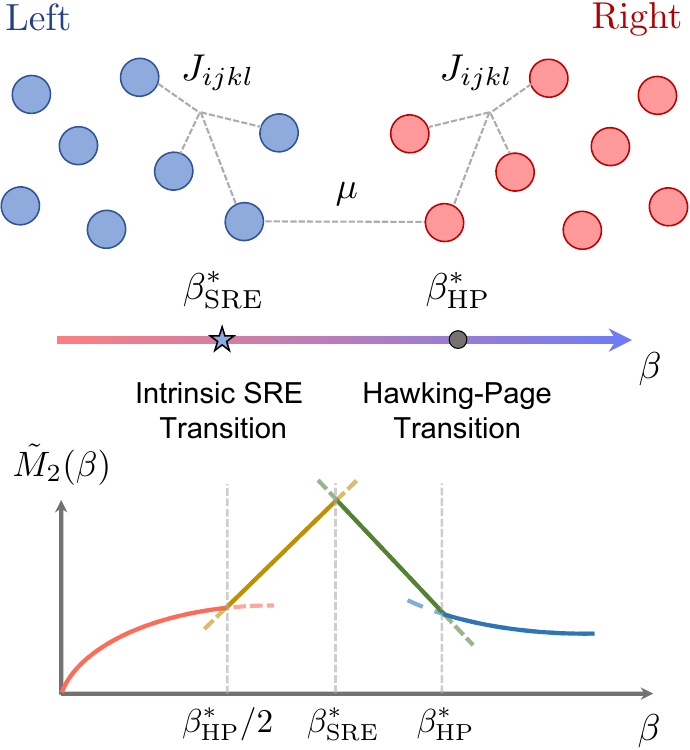}
    \caption{A schematic of our main results. The Maldacena-Qi coupled SYK model consists of two identical copies of the SYK model with the same random couplings $J_{ijkl}$ and a direct hopping $\mu$. In thermal equilibrium, the model exhibits a first-order Hawking-Page transition at $\beta^*_{\text{HP}}$ for $\mu\ll J$. We find that the (second) stabilizer R\'enyi entropy $\tilde{M}_2(\beta)$ exhibits its own intrinsic first-order transition at $\beta^*_{\text{SRE}}$, which is not reflected in thermodynamic quantities. Solid lines correspond to the dominant saddle, while dashed lines denote unstable saddles.}
    \label{fig:schemticas}
  \end{figure}

  In this Letter, we establish a general path-integral formalism for calculating the SRE in a class of solvable Sachdev-Ye-Kitaev (SYK) models~\cite{Sachdev:1992fk,kitaev2014talk,maldacena2016remarks,RevModPhys.94.035004}, which allows for exact analysis in the large-$N$ limit. We focus on the thermal ensemble of the Maldacena-Qi coupled SYK model~\cite{Maldacena:2018lmt,Qi:2020ian,Plugge:2020wgc} (illustrated in FIG.~\ref{fig:schemticas}), which exhibits a Hawking-Page transition from a low-temperature wormhole phase to a high-temperature black hole phase. We show that its SRE, while remaining extensive as the inverse temperature $\beta$ is tuned, exhibits three first-order transitions. Notably, one of these transitions cannot be detected by the system’s equilibrium properties and represents an intrinsic transition of the SRE~\cite{Ding:2025nua}. We explain that this transition arises from a change in connectivity between different branches of the path-integral contour, closely analogous to the celebrated replica wormholes~\cite{Penington:2019kki,Almheiri:2019qdq,Chen:2020wiq}. Our results pave the way for studying the SRE of strongly interacting fermions using large-$N$ solvable models and demonstrate that it can reveal hidden structures in many-body systems that are inaccessible to equilibrium observables, with potential connections to holography.

  \emph{ \color{blue} SRE \& Auxiliary Spins.--} To be concrete, we consider a quantum many-body system with $2N$ Majorana fermions, denoted by $\psi_{m}$ for $m \in \{1,2,\ldots,2N\}$. Following the standard convention in Ref.~\cite{maldacena2016remarks}, we adopt the canonical anticommutation relation $\{\psi_m,\psi_n\}=\delta_{mn}$. The definition of the SRE in Majorana fermion systems requires introducing an orthonormal operator basis, known as Majorana string operators
  \begin{equation}
  \Psi_{\bm{v}}=i^{|\bm{v}|(|\bm{v}|-1)/2}2^{|\bm{v}|/2}\psi^{v_1}_1\psi^{v_2}_2...\psi^{v_{2N}}_{2N}.
  \end{equation}
  Here, $\bm{v}$ is an $2N$-dimensional vector with elements $v_m \in \{0,1\}$, specifying the locations of the nontrivial Majorana operators and $|\bm{v}|\equiv \sum_m v_m$. The additional factors are chosen so that $\Psi_{\bm{v}}$ is Hermitian and satisfies $\Psi_{\bm{v}}^2 = 1$. This set of Majorana strings admits a one-to-one correspondence with Pauli string operators via the Jordan-Wigner transformation \cite{fradkin2013field}. A generic density matrix $\rho$ can be expanded in terms of Majorana strings
  \begin{equation}
  \rho=\frac{1}{2^N}\sum_{\bm{v}}\text{tr}[\rho \Psi_{\bm{v}}] \Psi_{\bm{v}}\equiv \frac{1}{2^N}\sum_{\bm{v}}c_{\bm{v}}(\rho)\Psi_{\bm{v}}.
  \end{equation}
  The coefficient $c_{\bm{v}}(\rho)$, referred to as the Majorana spectrum~\cite{Bera:2025pfp}, is real and satisfies the normalization condition $\sum_{\bm{v}} c_{\bm{v}}(\rho)^2 = 2^N e^{-S_2(\rho)}$, where the second R\'enyi entropy of the density matrix is defined as $S_2(\rho) = -\ln \text{tr}[\rho^2]$. Notice that due to the fermion parity, $c_{\bm{v}}(\rho)=0$ for arbitrary fermionic Majorana string with odd $|\bm{v}|$. 

   { A special class of quantum states, known as stabilizer states, admits an efficient classical representation. In fermionic systems \cite{Bera:2025pfp}, each pure stabilizer state is uniquely specified as the simultaneous +1 eigenstate of $N$ independent, mutually commuting Majorana strings $\{\pm\Psi_{\bm{v}_1},\pm\Psi_{\bm{v}_2},...,\pm\Psi_{\bm{v}_N}\}$, which serve as the generators of the stabilizer group. Consequently, $c_{\bm{v}}(\rho) = \pm1$ if $\Psi_{\bm{v}}$ is an element of the stabilizer group, and $c_{\bm{v}}(\rho) = 0$ otherwise. This observation motivates the introduction of the SRE $\tilde{M}_2(\rho)=M_2(\rho)-S_2(\rho)$ in Refs.~\cite{Leone:2021rzd,Haug:2023hcs,Wang:2023uog}, to quantify deviations from stabilizer states, i.e., the presence of quantum magic, with}
  \begin{equation}\label{eq:defM2}
  M_2(\rho)\equiv -\ln \left(2^{-N}\sum_{\bm{v}}c_{\bm{v}}(\rho)^4\right),
  \end{equation}
  which is a close analogy to the inverse participation number commonly used in the study of localization \cite{MacKinnon:1993wkx}. $\tilde{M}_2(\rho)$ vanishes if the density matrix is a pure stabilizer state or a maximally mixed state within the code space defined by a set of stabilizers.
  
  \begin{figure}[t]
    \centering
    \includegraphics[width=0.98\linewidth]{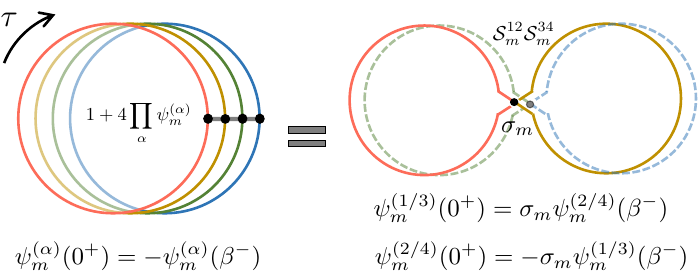}
    \caption{An illustration of the path-integral representation of the stabilizer R\'enyi entropy, either with extensive operator insertions (Eq.\eqref{eq:fourop}) or with fluctuating phases (Eq.\eqref{eq:final}). Lines of different colors correspond to different replicas of the density matrix.}
    \label{fig:pathintegral}
  \end{figure}

  We primarily focus on the path-integral representation of the SRE for thermal density matrices $\rho = e^{-\beta H}/Z_\beta$, associated with a Hamiltonian $H$, where the partition function is given by $Z_\beta = \text{tr}[e^{-\beta H}] \equiv e^{-S_\beta}$. In this case, we have $S_2(\rho) = S_{2\beta}-2S_\beta$. Therefore, it suffices to establish a path-integral formulation of $M_2(\rho)$. We consider $Z_{\text{SRE}}\equiv Z_\beta^4\sum_{\bm{v}}c_{\bm{v}}(\rho)^4 $, which reads \cite{Ding:2025nua}
  \begin{equation}
  Z_{\text{SRE}}=\sum_{\bm{v}}\text{tr}[({e^{-\beta H}})^{\otimes 4}\Psi_{\bm{v}}^{(1)}\Psi_{\bm{v}}^{(2)}\Psi_{\bm{v}}^{(3)}\Psi_{\bm{v}}^{(4)}].
  \end{equation}
  Here, we introduce four copies of the original system, labeled by $\alpha \in \{1,2,3,4\}$, with the corresponding Majorana strings denoted by $\Psi_{\bm{v}}^{(\alpha)}$. The ordering of these Majorana strings is irrelevant, since only bosonic ones contribute. Introducing $Z_{\text{SRE}}=e^{-S_{\text{SRE}}}$, we have the relation $M_2(\rho)=S_{\text{SRE}}-4S_\beta +N\ln 2$. Now, we rewrite the summation over Majorana strings as a summation on each site $m$. This yields
  \begin{equation}\label{eq:fourop}
  Z_{\text{SRE}}=\text{tr}\left[({e^{-\beta H}})^{\otimes 4}\prod_m\left(1+4\psi_{m}^{(1)}\psi_{m}^{(2)}\psi_{m}^{(3)}\psi_{m}^{(4)}\right)\right].
  \end{equation}
  Therefore, evaluating the SRE amounts to inserting operators on each site of the replicated system. When the density matrix is represented as an imaginary-time path integral~\cite{altland2010condensed}, these operators are inserted at a single imaginary time, chosen as $\tau = 0$. {A sketch is presented in Fig.~\ref{fig:pathintegral}, where each copy of the imaginary-time evolution $e^{-\beta H}$ is depicted as a closed circle, with different copies shown in distinct colors. The black dots denote operator insertions that couple the different copies. A similar replica trick has been widely used in calculations of entanglement entropy \cite{Calabrese:2009qy}. }

  To eliminate the extensive operator insertions, which are difficult to handle directly, we introduce a set of auxiliary Ising spins $\sigma_m = \pm 1$ and express $M_2(\rho)$ by
  \begin{equation}\label{eq:final}
  Z_{\text{SRE}}=\sum_{\{\sigma_m\}}\text{tr}\left[({e^{-\beta H}})^{\otimes 4}\prod_m \mathcal{S}_{m}^{12}[\sigma_m]\mathcal{S}_{m}^{34}[\sigma_m]\right],
  \end{equation}
  where $\mathcal{S}_{m}^{\alpha\beta}[\sigma_m]= e^{\frac{\pi}{2}\sigma_m\psi_m^{(\alpha)}\psi_m^{(\beta)}}$ is the fermionic SWAP operation that satisfies
  \begin{equation}\label{eq:bdy}
  \mathcal{S}_{m}^{\alpha\beta}[\sigma_m]^{\dagger}\begin{pmatrix}\psi^{(\alpha)}_m\\\psi^{(\beta)}_m\end{pmatrix}\mathcal{S}_{m}^{\alpha\beta}[\sigma_m]=\sigma_m\begin{pmatrix}\psi^{(\beta)}_m\\-\psi^{(\alpha)}_m\end{pmatrix}.
  \end{equation}
  Without these SWAP operations, the four branches of the path integral remain disconnected, each obeying the anti-periodic boundary condition for fermionic fields. The SWAP operations effectively connect each pair of branches, introducing a fluctuating phase of $\pm 1$ when fields pass through the intersections, as illustrated in Fig.~\ref{fig:pathintegral}. The relative minus sign on the R.H.S. ensures the standard anti-periodic boundary condition after two branches are joined. Eq.~\eqref{eq:final} thus provides a geometric interpretation of the SRE for generic systems. A similar expression for qubit systems are presented in the supplementary material~\cite{SM}. 

  \emph{ \color{blue} Coupled SYK model.--} For generic many-body systems, the path integral in Eq.\eqref{eq:final} involves $2N$ independent Ising spins, whose evaluation requires numerical simulations using techniques such as quantum Monte Carlo~\cite{Ding:2025nua}. In contrast, we are interested in analyzing the SRE in solvable SYK models~\cite{Sachdev:1992fk,kitaev2014talk,maldacena2016remarks,Kitaev:2017awl}. Numerical calculations of the SRE in the SYK model have been performed in \cite{Bera:2025pfp,Jasser:2025myz,njgn-fksh} for finite $N$. Here, we primarily focus on the large-$N$ limit. We focus on the Maldacena-Qi coupled SYK model~\cite{Maldacena:2018lmt,Qi:2020ian,Plugge:2020wgc}, where the Majorana fermions are divided into two clusters, labeled left and right, with indices $m=L,j$ or $R,j$ for $j\in\{1,2,\ldots,N\}$. The Hamiltonian, illustrated in Fig.~\ref{fig:schemticas}, is given by 
  \begin{equation}
  \begin{aligned}
  H=&\sum_{i<j<k<l}J_{ijkl}\Big(\psi_{L,i}\psi_{L,j}\psi_{L,k}\psi_{L,l}\\
  &+\psi_{R,i}\psi_{R,j}\psi_{R,k}\psi_{R,l}\Big)+i\mu \sum_j \psi_{j,L}\psi_{j,R},
  \end{aligned}
  \end{equation}
  where the random four-fermion couplings are independent Gaussian variables with 
  \begin{equation}
  \overline{J_{ijkl}}=0,\ \ \ \ \ \overline{J_{ijkl}^2}={6J^2}/{N^3}.
  \end{equation}
  At $\mu=0$, the model describes two decoupled SYK systems, whose low-energy sector is dual to an AdS$_2$ black hole~\cite{Maldacena:2016upp}. A weak coupling ($\mu \ll J$) becomes relevant in the low-temperature limit, driving a first-order Hawking-Page transition at $\beta^*_{\text{HP}}$ to an eternal wormhole~\cite{Maldacena:2018lmt}. For $\mu/J \gtrsim 0.12$, this transition turns into a crossover, while showing qualitatively the same features.

  The saddle-point equation for the path integral~\eqref{eq:final} can be derived by generalizing the standard SYK techniques for equilibrium properties~\cite{maldacena2016remarks}. The central idea is that, after disorder averaging, pairs of modes $L,j$ and $R,j$ with different $j$ become equivalent, allowing the system to be analyzed by focusing on just two representative modes with fixed $j$. The effects of all other modes are incorporated solely through a self-consistent self-energy term. Leaving the detailed derivation to the supplementary material~\cite{SM}, the saddle-point equations read
  \begin{equation}\label{eq:saddle}
  \begin{aligned}
  &\Sigma_{ss'}^{(\alpha\beta)}(\tau,\tau')=J^2G_{ss'}^{(\alpha\beta)}(\tau,\tau')^3,\\
  &G_{ss'}^{(\alpha\beta)}=\frac{\sum_{\sigma_L,\sigma_R}g_{ss',\sigma_L\sigma_R}^{(\alpha\beta)} \text{det}[g_{ss',\sigma_L\sigma_R}^{(\alpha\beta)}]^{-1/2}}{\sum_{\sigma_L,\sigma_R} \text{det}[g_{ss',\sigma_L\sigma_R}^{(\alpha\beta)}]^{-1/2}}.
  \end{aligned}
  \end{equation}
  Here, $G_{ss'}^{(\alpha\beta)}(\tau,\tau') = \big\langle \psi_{s}^{(\alpha)}(\tau)\psi_{s'}^{(\beta)}(\tau') \big\rangle$ denotes the full Green’s function with $s,s' \in \{L,R\}$, and $\Sigma_{ss'}^{(\alpha\beta)}(\tau,\tau')$ is the corresponding self-energy. For conciseness, the index $j$ has been suppressed. We also introduce $g_{ss',\sigma_L\sigma_R}^{(\alpha\beta)}$ to denote the Green’s function evaluated with the fixed boundary condition~\eqref{eq:bdy}:
  \begin{equation}
  g_{ss',\sigma_L\sigma_R}^{(\alpha\beta)}=\left(\partial_\tau-\mu(\sigma_y)_{ss'}-\Sigma_{ss'}^{(\alpha\beta)}\right)^{-1}_{\sigma_L\sigma_R}.
  \end{equation}
  The subscript $\sigma_{L/R}=\pm 1$ specifies the boundary condition of the corresponding mode, which in turn determines the definition of the kinetic term $\partial_\tau$. The inversion is carried out by treating $g$ as a matrix in the space of $s$, $\alpha$, and $\tau$ for fixed $\sigma_L$ and $\sigma_R$. After solving the saddle-point equations, the action $S_{\text{SRE}}$ is given by
  \begin{equation}
  \begin{aligned}
  \frac{S_{\text{SRE}}}{N}=&-\ln\Bigg[\sum_{\sigma_L,\sigma_R} \text{det}[g_{ss',\sigma_L\sigma_R}^{(\alpha\beta)}]^{-1/2}\Bigg]\\
  &+\frac{3}{8}\sum_{\alpha\beta,ss'}\int_{\tau,\tau'} G_{ss'}^{(\alpha\beta)}(\tau,\tau')\Sigma_{ss'}^{(\alpha\beta)}(\tau,\tau').
  \end{aligned}
  \end{equation}
  Together with the thermal partition function $Z_\beta$~\cite{maldacena2016remarks}, this determines $M_2$ and hence the SRE $\tilde{M}_2$.

  We further discuss the symmetry of the path integral. It is invariant under $\sigma_{L/R} \rightarrow -\sigma_{L/R}$ and
  \begin{equation}
  \big(\psi_{L/R}^{(\alpha_1)},\psi_{L/R}^{(\alpha_2)}\big)\rightarrow -\big(\psi_{L/R}^{(\alpha_1)},\psi_{L/R}^{(\alpha_2)}\big),
  \end{equation}
  for chosen $\alpha_1\in\{1,2\}$ and $\alpha_2\in\{3,4\}$. Assuming this symmetry is not spontaneously broken, as verified numerically, the full Green’s function $G_{ss'}^{(\alpha\beta)}$ is expected to be diagonal in $\alpha\beta$. Moreover, permutation symmetry implies that $G_{ss'}^{(\alpha\alpha)}$ is independent of $\alpha$. This results in $G_{ss'}^{(\alpha\beta)}=\delta^{\alpha\beta}G_{ss'}$, $\Sigma_{ss'}^{(\alpha\beta)}=\delta^{\alpha\beta}\Sigma_{ss'}$. In addition, we have
  \begin{equation}\label{eqn:symmetry}
  g_{ss',\sigma_L\sigma_R}^{(\alpha\beta)}=f^{\alpha}f^{\beta}g_{ss',-\sigma_L-\sigma_R}^{(\alpha\beta)},
  \end{equation}
  where $f^\alpha=-1$ if $\alpha=\alpha_1$ or $\alpha_2$ and otherwise $f^\alpha=1$. This also leads to $\text{det}[g_{ss',\sigma_L\sigma_R}^{(\alpha\beta)}]=\text{det}[g_{ss',-\sigma_L-\sigma_R}^{(\alpha\beta)}]$.
 
  \begin{figure}[t]
    \centering
    \includegraphics[width=0.98\linewidth]{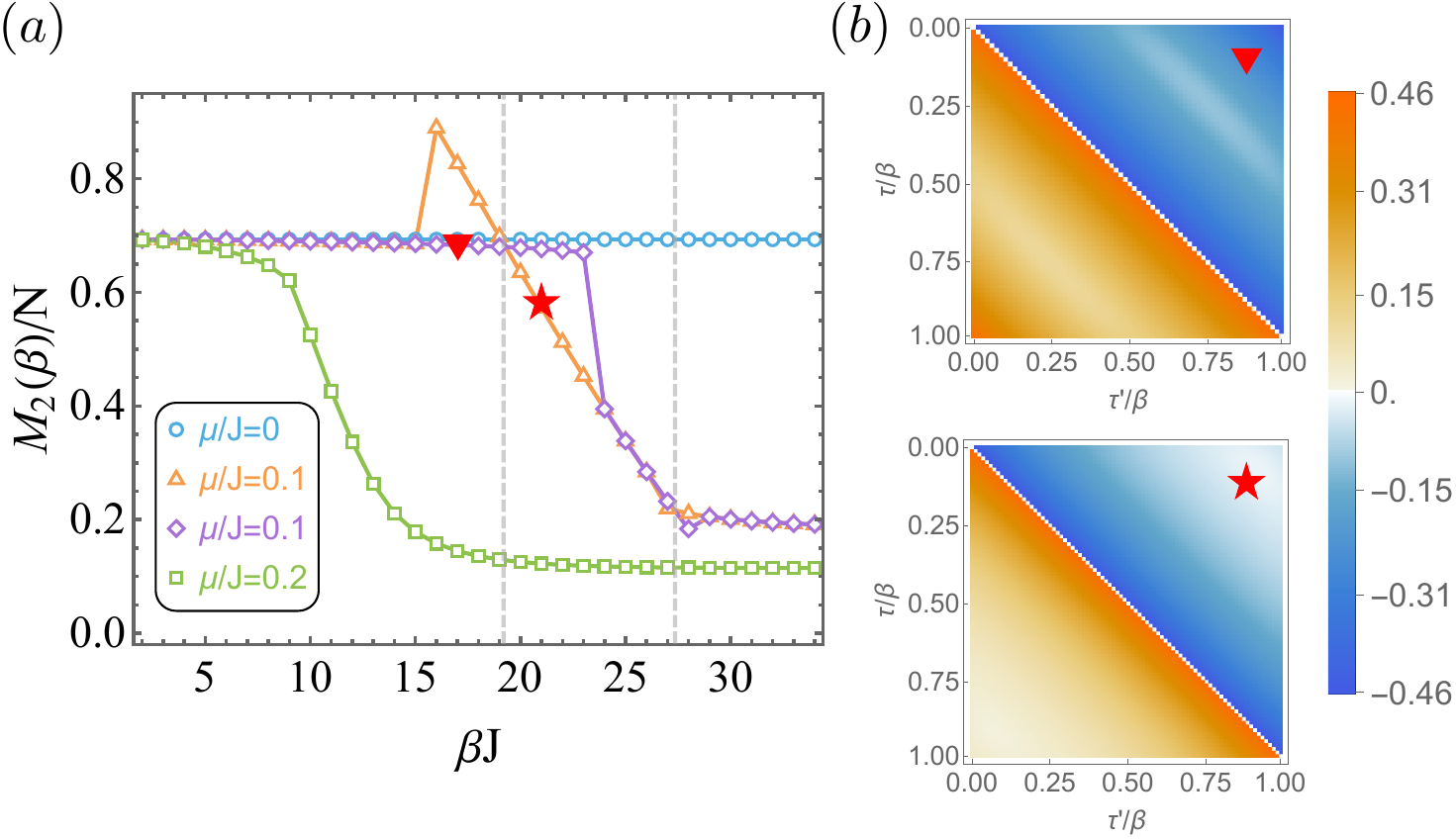}
    \caption{(a) Numerical results for $M_2(\beta)$, obtained by iteratively solving the saddle-point equations \eqref{eq:saddle} for $\mu/J \in {0,0.1,0.2}$. The orange triangles and purple diamonds correspond to solutions with different initializations, favoring the high- and low-temperature solutions, respectively. (b) The Green's function $G^{(11)}_{LL}(\tau,\tau')$ is plotted at $\beta J=17$ and $\beta J=21$ for $\mu/J=0.1$, clearly exhibiting qualitatively different features near $\tau=0^+$ and $\tau'=\beta^-$.  }
    \label{fig:numerics}
  \end{figure}

  \emph{ \color{blue} Results.--} The analytical solution of \eqref{eq:saddle} is challenging due to the breaking of time-translation symmetry induced by the auxiliary spin $\sigma_s$. Therefore, we perform a numerical study by discretizing the imaginary-time evolution and solving the Green’s function iteratively. Similar methods have been applied previously to the study of the R\'enyi entropy in SYK models~\cite{Chen:2020wiq,Zhang:2020kia,Zhang:2022yaw}. The results for $M_2$ as a function of the inverse temperature $\beta$ are presented in Fig.~3 for $\mu/J = 0$, $0.1$, and $0.2$. We first focus on the case $\mu/J = 0$, where the system reduces to two decoupled SYK models. Numerical results suggest that $M_2(\beta) = N \ln 2$ is independent of temperature, indicating that only the identity operator contributes in Eq.~\eqref{eq:defM2} at $N\rightarrow \infty$. To provide an analytical derivation of this relation, we note that since the two sides are uncorrelated, it suffices to focus on $g^{(\alpha\beta)}=g^{(\alpha\beta)}_{LL,11}$ and $\Sigma^{(\alpha\beta)}=\Sigma^{(\alpha\beta)}_{LL}$. Taking into account the symmetry \eqref{eqn:symmetry}, the saddle-point equations become
  \begin{equation}
  \begin{aligned}
  &\Sigma^{(\alpha\beta)}(\tau,\tau')=J^2\delta^{\alpha\beta}g^{(\alpha\beta)}(\tau,\tau')^3,
  \end{aligned}
  \end{equation}
  with $g^{(\alpha\beta)}=\left(\partial_\tau-\Sigma^{(\alpha\beta)}\right)^{-1}_{\sigma=1}$. This coincides with the saddle-point equation for the norm of the Kourkoulou-Maldacena (KM) state~\cite{Kourkoulou:2017zaj,Zhang:2020iep}, and $S_{\text{SRE}}$ corresponds to the action of the KM state, $8S_{\text{KM}}$, with the overall factor arising from the replicas and flavors. On the other hand, it has been established that $S_{\text{KM}} + C_0= S_{\beta,\text{SYK}} = S_\beta/2 $, where $S_{\beta,\text{SYK}}$ is the action of a single SYK model and $C_0$ is independent of $\beta J$~\cite{Kourkoulou:2017zaj,Zhang:2020iep}. Therefore, we conclude that $S_{\text{SRE}}-4S_\beta$ is independent of $\beta J$. Its value can then be fixed by evaluating at $\beta J=0$, which yields $M_2(\beta) = N \ln 2$. Similar conclusions hold for other SYK-like models with permutation symmetry, including the non-chaotic SYK$_2$ model and those with mixed SYK$_2$ and SYK$_4$ interactions. An alternative derivation is provided in the supplementary material~\cite{SM}.

  \begin{figure}[t]
    \centering
    \includegraphics[width=0.98\linewidth]{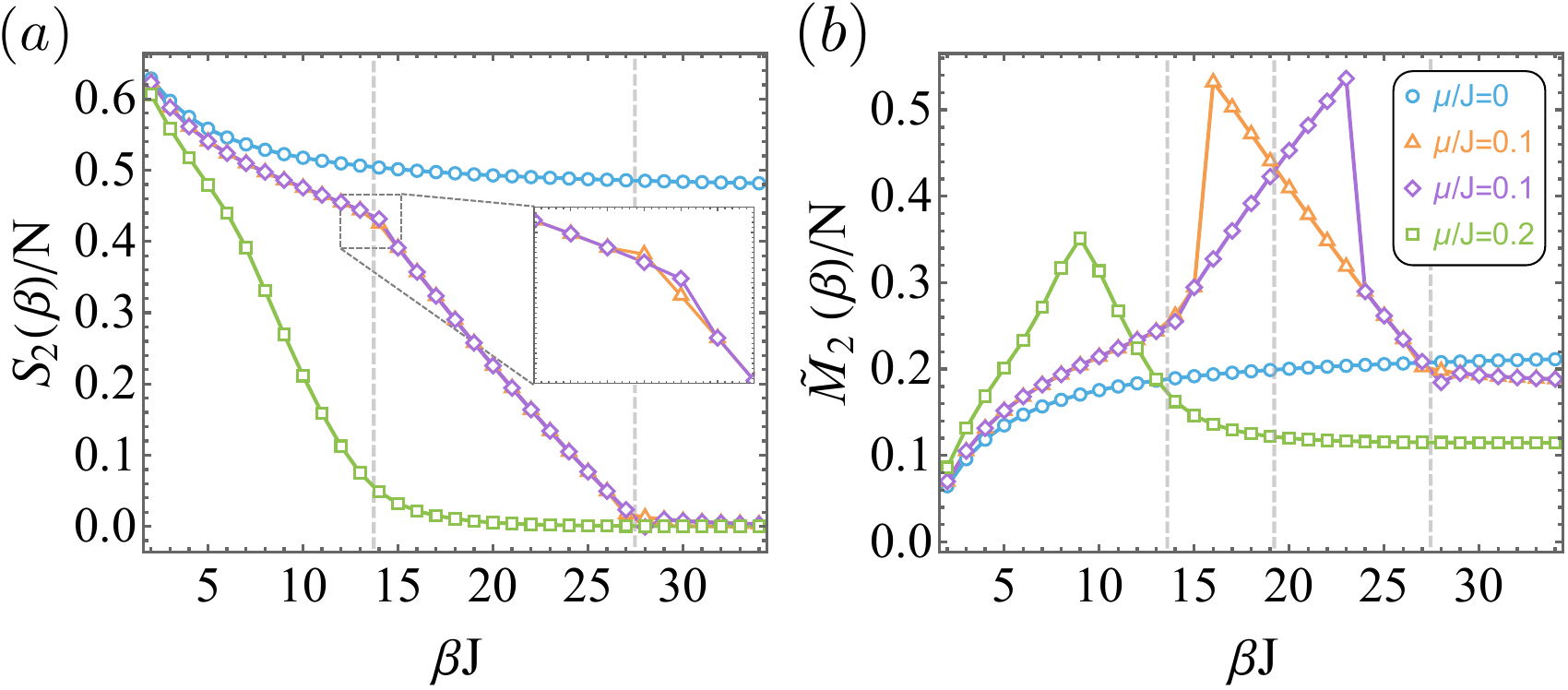}
    \caption{Numerical results for (a) $S_2(\beta)$, obtained by solving the saddle-point equations for the partition function~\cite{Maldacena:2018lmt}, and (b) $\tilde{M}_2(\beta)$, computed using $M_2(\beta)$ from FIG.~\ref{fig:numerics}. The results reveal three first-order transitions of the SRE for $\mu/J=0.1$, occurring at $\beta^*_{\text{HP}}/2$, $\beta^*_{\text{SRE}}$, and $\beta^*_{\text{HP}}$. }
    \label{fig:numerics2}
  \end{figure}
  
  Next, we introduce a weak coupling between the two sides with $\mu/J=0.1$. In the high-temperature limit, the coupling effect is perturbative, and $M_2(\beta) \approx N\big(\ln 2 - O(\mu^2 \beta^2)\big)$. As the inverse temperature increases, we observe a first-order transition of $S_{\text{SRE}}$ near $\beta^*_{\text{SRE}} J \approx 19$, evidenced by the coexistence of two distinct solutions to Eq.\eqref{eq:saddle}, obtained from different initialization conditions in the iteration. The physical solution is the one with smaller $M_2(\beta)$ and the SRE becomes singular. Similar singularities have been observed in quantum Monte Carlo simulations of quantum spin models \cite{Ding:2025nua}. Fig.\ref{fig:numerics}(b) shows the Green's function $G^{(11)}_{LL}(\tau,\tau')$. The solution corresponding to the high-temperature saddle resembles the thermal Green's function, approaching $1/2$ at $\tau=0^+$ and $\tau'=\beta^-$. This behavior indicates that the SWAP operators are effectively disrupted by the fluctuating field $\sigma$, leading to a disconnection between replicas. In contrast, the Green's function at low temperatures decays as $\tau - \tau'$ increases, consistent with full connectivity between replicas (1) and (2), where $\tau=0^+$ and $\tau'=\beta^-$ correspond to maximal imaginary-time separation. Therefore, the transition reflects a qualitative change in replica connectivity. {From the perspective of fermionic modes, the coupling between the two sides becomes dominant in the low-temperature saddle. This drives the system toward a target stabilizer state with stabilizers $\{-i\psi_{j,L}\psi_{j,R}\}$, thereby leading to a reduction in $M_2(\beta)$. }

  If we further increase $\beta$, another first-order transition appears near $\beta^*_{\text{HP}} J \approx 27$. This is the Hawking-Page transition of the thermal partition function, originating from the contribution of $S_\beta$ in the relation $M_2(\rho) = S_{\text{SRE}} - 4S_\beta+N\ln 2$. The physical solution corresponds to the branch with smaller $S_\beta$, and therefore larger $M_2(\beta)$. For $\beta J \rightarrow \infty$, $M_2(\beta)$ saturates to a value significantly smaller than in the case with $\mu=0$. In the supplementary material~\cite{SM}, we further show that $M_2(\infty) = N(2\ln 2 - 4s_0) < N\ln 2$ for $\mu/J \rightarrow 0$, where $s_0 \approx 0.2324$ is the zero-temperature entropy density of the SYK$_4$ model~\cite{maldacena2016remarks}. Consequently, $M_2(\infty)$ is discontinuous at $\mu=0$ as $\mu/J$ is varied. For larger $\mu/J = 0.2$, the transitions of $M_2(\beta)$ turn into crossovers, while the qualitative behavior remains unchanged.

  Finally, we combine $M_2(\beta)$ with the second R\'enyi entropy $S_2(\beta)$ to obtain the SRE $\tilde{M}_2(\beta)$. The results are shown in FIG.~\ref{fig:numerics2}. For the $\mu/J=0$ case, $S_2(\beta)$ decreases smoothly from $\ln 2$ to $2s_0$ as the inverse temperature increases, leading to a monotonic increase of the SRE. When a weak coupling $\mu/J$ is introduced, the second R\'enyi entropy $S_2(\beta) = S_{2\beta} - 2S_\beta$ exhibits first-order transitions at both $\beta^*_{\text{HP}}/2$ and $\beta^*_{\text{HP}}$. In particular, the entropy decreases almost linearly in the intermediate regime and saturates near zero for $\beta > \beta^*_{\text{HP}}$. Numerically, we find that $\beta^*_{\text{HP}}<2\beta^*_{\text{SRE}}$. Therefore, the SRE exhibits a linear increase for $\beta \in (\beta^*_{\text{HP}}/2,\beta^*_{\text{SRE}})$. The increase terminates due to the intrinsic transition at $\beta^*_{\text{SRE}}$, and the SRE then decreases and saturates through another transition at $\beta^*_{\text{HP}}$. This non-monotonic behavior with singularities resembles the entanglement barrier \cite{Dubail:2016xht,Wang:2019ued,PhysRevB.104.014301}, which represents an obstacle for classical simulations using tensor network approaches. Intuitively, our results suggest a corresponding barrier in the classical resources required to simulate the model in the intermediate temperature regime using stabilizer-based methods.

  \emph{ \color{blue} Discussions.--}
  In this Letter, we develop a general path-integral framework for computing the stabilizer R\'enyi entropy in solvable SYK models, providing a tool to study the quantum magic of strongly interacting fermions in the thermodynamic limit. Applying this framework to the Maldacena-Qi model, we uncovered multiple first-order transitions of the SRE, including an intrinsic transition that is invisible to conventional thermodynamic observables. Our results demonstrate that the SRE captures features of many-body quantum states beyond standard equilibrium measures, revealing a new type of phase behavior that is inherently informational. Moreover, the non-monotonic temperature dependence of the SRE reveals a temperature-dependent barrier to the resources required for classical simulation.

  We conclude with a few remarks on future directions. First, the path-integral representation introduced here, employing auxiliary spins and fermionic SWAP operations, provides a versatile tool that can be applied to other interacting systems, including higher-dimensional SYK models and systems undergoing non-equilibrium dynamics. Second, it would be interesting to study the transitions of the stabilizer R\'enyi entropy from a holographic perspective and explore possible connections to replica wormholes, whose SYK counterparts also arise from first-order transitions with distinct Green’s function structures~\cite{Chen:2020wiq}. Finally, it remains an open question whether similar transitions appear in other measures of quantum magic, such as mana. We leave these explorations for future work.

\vspace{5pt}
\textit{Acknowledgement.} 
We thank Langxuan Chen, Yingfei Gu, Haolin Jiang, Zeyu Liu, Yuzhi Tong, and Zhao-Yi Zeng for helpful discussions. This project is supported by the NSFC under grant 12374477, the Shanghai Rising-Star Program under grant number 24QA2700300, and the Quantum Science and Technology-National Science and Technology Major Project 2024ZD0300101.

\bibliography{main.bbl}

\ifarXiv


\section*{Supplementary material}

# Supplementary Material: Stabilizer Rényi Entropy and its Transition in the Coupled Sachdev-Ye-Kitaev Model

Pengfei Zhang,[1, 2, *] Shuyan Zhou,[1, 3] and Ning Sun[1]

[1]*Department of Physics & State Key Laboratory of Surface Physics, Fudan University, Shanghai, 200438, China*
[2]*Hefei National Laboratory, Hefei 230088, China*
[3]*Department of Physics & Center for Theory of Quantum Matter,*
*University of Colorado Boulder, Boulder, CO 80309, USA*

(Dated: September 22, 2025)

In this supplementary material, we present (1) the representation of the SRE using auxiliary fields for qubit systems, (2) the derivation of the saddle-point equations, (3) an alternative proof of $M_2(\beta) = N \ln 2/2$ for the single-site SYK model, and (4) the analysis of $M_2(\beta)$ in the low-temperature limit.

## THE REPRESENTATION OF THE SRE USING AUXILIARY FIELDS FOR QUBIT SYSTEMS

In the main text, we focus on systems composed of Majorana fermions and derive a representation of the SRE using auxiliary fields. In this section, we show that a similar representation also applies to qubit systems. Specifically, we consider quantum systems with $N$ qubits, where the single-site Pauli operators (including identity operators) are denoted by $\sigma_m^\eta$ with $\eta \in \{0, 1, 2, 3\}$. A natural operator basis is given by the Pauli string operators:

$$P_{\boldsymbol{\eta}} = \sigma_1^{\eta_1} \sigma_2^{\eta_2} ... \sigma_N^{\eta_N}. \tag{1}$$

Here, $\boldsymbol{\eta} = (\eta_1, \eta_2, ..., \eta_N)$ is a $N$-dimensional vector. Expanding the density matrix $\rho$ in $P_{\boldsymbol{\eta}}$, we find

$$\rho = \frac{1}{2^N} \sum_{\boldsymbol{\eta}} \text{tr}[\rho P_{\boldsymbol{\eta}}] P_{\boldsymbol{\eta}} \equiv \frac{1}{2^N} \sum_{\boldsymbol{\eta}} c_{\boldsymbol{\eta}}(\rho) P_{\boldsymbol{\eta}}. \tag{2}$$

The SRE of the qubit system is then defined as $\tilde{M}_2(\rho) = M_2(\rho) - S_2(\rho)$ [1–3], with

$$M_2(\rho) \equiv -\ln \left( 2^{-N} \sum_{\boldsymbol{\eta}} c_{\boldsymbol{\eta}}(\rho)^4 \right), \tag{3}$$

To derive an alternative representation with auxiliary fields, analogous to the Majorana case, we introduce four replicas of the original system and express

$$2^N e^{-M_2} = \sum_{\boldsymbol{\eta}} c_{\boldsymbol{\eta}}(\rho)^4 = \sum_{\boldsymbol{\eta}} \text{tr}[\rho^{\otimes 4} P_{\boldsymbol{\eta}}^{(1)} P_{\boldsymbol{\eta}}^{(2)} P_{\boldsymbol{\eta}}^{(3)} P_{\boldsymbol{\eta}}^{(4)}]. \tag{4}$$

Here, $P_{\boldsymbol{\eta}}^{(\alpha)}$ denotes the Pauli string operator acting on the $\alpha$-th replica. We then re-express the formula as a sum over Pauli operators on each individual site, yielding

$$\begin{aligned}
\sum_{\boldsymbol{\eta}} c_{\boldsymbol{\eta}}(\rho)^4 &= \frac{1}{2^N} \text{tr}\left[ \rho^{\otimes 4} \prod_{m,\eta} \left( 1 + \sigma_m^{\eta,(1)} \sigma_m^{\eta,(2)} \sigma_m^{\eta,(3)} \sigma_m^{\eta,(4)} \right) \right] \\
&= \frac{1}{2^N} \sum_{\{s_m^\eta\}} \text{tr}\left[ \rho^{\otimes 4} \prod_{m,\eta} e^{i\frac{\pi}{4}\sigma_m^{\eta,(1)}\sigma_m^{\eta,(2)}s_m^\eta} e^{-i\frac{\pi}{4}\sigma_m^{\eta,(3)}\sigma_m^{\eta,(4)}s_m^\eta} \right] \\
&= \frac{1}{2^N} \sum_{\{s_m^\eta\}} \text{tr}\left[ \rho^{\otimes 4} \prod_m \mathcal{S}_m^{(12)}(s_m^\eta) \mathcal{S}_m^{(34)}(-s_m^\eta) \right],
\end{aligned} \tag{5}$$

where we introduce the Ising variables $s_m^\eta = \pm 1$ and phase-dependent SWAP operator $\mathcal{S}_m^{(12)}(s_m^\eta) = \prod_\eta e^{i\frac{\pi}{4}\sigma_m^{\eta,(1)}\sigma_m^{\eta,(2)}s_m^\eta}$, which satisfies

$$\mathcal{S}_m^{(12)}(s_m^\eta)^\dagger \begin{pmatrix} \sigma_m^{x,(1)} \\ \sigma_m^{y,(1)} \\ \sigma_m^{z,(1)} \end{pmatrix} \mathcal{S}_m^{(12)}(s_m^\eta) = \begin{pmatrix} s_m^y s_m^z \sigma_m^{x,(2)} \\ s_m^x s_m^z \sigma_m^{y,(2)} \\ s_m^x s_m^y \sigma_m^{z,(2)} \end{pmatrix}. \tag{6}$$

This is analogous to Eq.(6) in the main text.

## DERIVATION OF THE SADDLE-POINT EQUATIONS

We now provide a detailed derivation of the path-integral representation and the saddle-point equations for the coupled SYK model. To begin, we rewrite Eq. (6) in the form of a path integral

$$Z_{\rm SRE} = \sum_{\{\sigma_{s,j}\}} \int dJ_{ijkl} P(J_{ijkl}) \int_{\sigma_{s,j}} D\psi_{s,j}^{(\alpha)} \, \exp\Big(-S[\psi_{s,j}^{(\alpha)}]\Big),$$
$$S = \int_0^\beta d\tau \left[ \sum_{s,j,\alpha} \frac{1}{2} \psi_{s,j}^{(\alpha)} \partial_\tau \psi_{s,j}^{(\alpha)} + \sum_\alpha H[\psi_{s,j}^{(\alpha)}] \right]. \tag{7}$$

Here, the subscript $\sigma_{s,j}$ denotes the boundary condition specified in Fig. 2. Using the self-averaging property of the SYK model in the large-$N$ limit [4], we perform the disorder average directly on $Z_{\rm SRE}$. Since the integration over $J_{ijkl}$ is Gaussian, it can be carried out explicitly, yielding

$$Z_{\rm SRE} = \sum_{\{\sigma_{s,j}\}} \int_{\sigma_{s,j}} D\psi_{s,j}^{(\alpha)} \, \exp\bigg( - \sum_{ss',j,\alpha} \int d\tau \frac{1}{2} \psi_{s,j}^{(\alpha)}(\tau)[\delta_{ss'}\partial_\tau - \mu(\sigma_y)_{ss'}]\psi_{s',j}^{(\alpha)}(\tau) + \frac{J^2}{8N^3} \sum_{ss',\alpha\beta} \int d\tau d\tau' \Big[ \sum_j \psi_{s,j}^{(\alpha)}(\tau)\psi_{s',j}^{(\beta)}(\tau') \Big]^4 \bigg). \tag{8}$$

Next, we introduce the bilocal field $G_{ss'}^{(\alpha\beta)}(\tau,\tau') = \sum_j \psi_{s,j}^{(\alpha)}(\tau)\psi_{s',j}^{(\beta)}(\tau')/N$ together with the corresponding Lagrange multiplier $\Sigma_{ss'}^{(\alpha\beta)}(\tau,\tau')$. This leads to

$$Z_{\rm SRE} = \sum_{\{\sigma_{s,j}\}} \int_{\sigma_{s,j}} D\psi_{s,j}^{(\alpha)} DG_{ss'}^{(\alpha\beta)} D\Sigma_{ss'}^{(\alpha\beta)} \, \exp(-S_{G\Sigma}),$$
$$S_{G\Sigma} = \int d\tau d\tau' \bigg( \frac{1}{2} \psi_{s,j}^{(\alpha)}(\tau)[\hat{G}_0^{-1} - \Sigma_{ss'}^{(\alpha\beta)}(\tau,\tau')]\psi_{s',j}^{(\beta)}(\tau') + \frac{N}{2} G_{ss'}^{(\alpha\beta)}(\tau,\tau')\Sigma_{ss'}^{(\alpha\beta)}(\tau,\tau') - \frac{J^2 N}{8} G_{ss'}^{(\alpha\beta)}(\tau,\tau')^4 \bigg). \tag{9}$$

where we define $\hat{G}_0^{-1} = \delta^{\alpha\beta}\delta(\tau-\tau')[\delta_{ss'}\partial_\tau - \mu(\sigma_y)_{ss'}]$, with the summation over indices kept implicit. Importantly, after introducing the bilocal fields, modes with different $j$ completely decouple, and the integration over the Majorana fermions $\psi_{s,j}^{(\alpha)}$ can be performed as

$$Z_{\rm SRE} = \int DG_{ss'}^{(\alpha\beta)} D\Sigma_{ss'}^{(\alpha\beta)} \left[ \sum_{\sigma_L \sigma_R} \sqrt{\det[\hat{G}_0^{-1} - \Sigma]_{\sigma_L \sigma_R}} \right]^N$$
$$\exp\bigg( - \int d\tau d\tau' \left[ \frac{N}{2} G_{ss'}^{(\alpha\beta)}(\tau,\tau')\Sigma_{ss'}^{(\alpha\beta)}(\tau,\tau') - \frac{J^2 N}{8} G_{ss'}^{(\alpha\beta)}(\tau,\tau')^4 \right] \bigg). \tag{10}$$

Finally, in the limit $N \to \infty$, the integration over the bilocal fields is dominated by their saddle points. Varying with respect to the bilocal fields yields the saddle-point equations, given by Eqs. (10) and (11), while the action can be computed using Eq. (12).

## AN ALTERNATIVE DERIVATION OF THE $M_2$ FOR THE SINGLE-SITE SYK MODEL

In the main text, we proved that $M_2(\beta) = N \ln 2$ for the coupled SYK model with $\mu = 0$, which is equivalent to $M_2(\beta) = \frac{N}{2} \ln 2$ for a single SYK model with $N$ Majorana fermions. Here, we provide an alternative proof by directly showing that only the identity operator contributes. For a single SYK model, we have

$$Z_{\rm SRE} = {\rm tr} \left[ \left(e^{-\beta H_{\rm SYK}}\right)^{\otimes 4} \prod_j \left(1 + 4\psi_j^{(1)}\psi_j^{(2)}\psi_j^{(3)}\psi_j^{(4)}\right) \right]. \tag{11}$$

Next, we directly move to the path-integral representation, without introducing the auxiliary Ising fields. After introducing the bilocal fields, this leads to

$$Z_{\text{SRE}} = \int D\psi_j^{(\alpha)} DG^{(\alpha\beta)} D\Sigma^{(\alpha\beta)} \prod_j \left(1 + 4\psi_j^{(1)}(0)\psi_j^{(2)}(0)\psi_j^{(3)}(0)\psi_j^{(4)}(0)\right) e^{-S_0},$$

$$S_0 = \int_{\tau\tau'} \left(\frac{1}{2}\psi_j^{(\alpha)}(\tau)[\delta^{\alpha\beta}\delta_{\tau\tau'}\partial_{\tau'} - \Sigma^{(\alpha\beta)}(\tau,\tau')]\psi_j^{(\beta)}(\tau') + \frac{N}{2}G^{(\alpha\beta)}(\tau,\tau')\Sigma^{(\alpha\beta)}(\tau,\tau') - \frac{J^2 N}{8}G^{(\alpha\beta)}(\tau,\tau')^4\right). \quad (12)$$

Once again, modes with different $j$ completely decouple, and the integration over each Majorana fermion can be separated out. This yields

$$Z_{\text{SRE}} = \int DG^{(\alpha\beta)} D\Sigma^{(\alpha\beta)} \left[\int D\psi^{(\alpha)} \left(1 + 4\prod_\alpha \psi^{(\alpha)}(0)\right) e^{-\frac{1}{2}\int_{\tau\tau'} \psi^{(\alpha)}[\delta^{\alpha\beta}\delta_{\tau\tau'}\partial_{\tau'} - \Sigma^{(\alpha\beta)}]\psi^{(\beta)}}\right]^N e^{-\Delta S_{G\Sigma}}. \quad (13)$$

Here, $\Delta S_{G\Sigma} = \int_{\tau\tau'} \left[\frac{N}{2}G^{(\alpha\beta)}\Sigma^{(\alpha\beta)} - \frac{J^2 N}{8}(G^{(\alpha\beta)})^4\right]$. Next, by analogy with the symmetry of the coupled SYK model, the saddle-point solutions for both $G^{(\alpha\beta)}(\tau,\tau') = G(\tau,\tau')\delta^{(\alpha\beta)}$ and $\Sigma^{(\alpha\beta)}(\tau,\tau') = \Sigma(\tau,\tau')\delta^{(\alpha\beta)}$ are expected to be diagonal in $\alpha\beta$. Consequently, we expect that at the saddle-point, the contribution of $\prod_\alpha \psi^{(\alpha)}(0)$, computed by using Wick's theorem, vanishes. Imposing this constraint, the saddle-point equation of the SRE coincides with (four copies of) the thermal saddle-point equation, and the solution $G(\tau,\tau')$ is precisely the thermal two-point function. Therefore, we obtain

$$\frac{S_{\text{SRE}}}{N} = 4 \times \left(-\frac{1}{2}\ln\det(\partial_\tau - \Sigma) + \frac{3}{8}\int d\tau d\tau' G(\tau,\tau')\Sigma(\tau,\tau')\right) = 4S_{\beta,\text{SYK}}. \quad (14)$$

Finally, using the relation $M_2(\beta) = S_{\text{SRE}} - 4S_{\beta,\text{SYK}} + \frac{N}{2}\ln 2$ for a system with $N$ Majorana fermions, we obtain the final result $M_2(\beta) = \frac{N}{2}\ln 2$.

## THE ANALYSIS OF $M_2$ IN THE LOW-TEMPERATURE LIMIT

Here, we analyze $M_2(\beta)$ in the wormhole phase. As established in Ref. [5], for $\mu/J \ll 1$, the ground state of the coupled SYK model coincides with the thermofield double (TFD) state:

$$|\text{TFD}\rangle = \frac{1}{\sqrt{Z_{\beta_{\text{eff}},\text{SYK}}}} e^{-\beta_{\text{eff}} H_{\text{SYK}}/2}|\text{EPR}\rangle. \quad (15)$$

Here, $|\text{EPR}\rangle$ is the ground state of $H$ at $\mu \to \infty$, which is a maximally entangled state between two sides. $H_{\text{SYK}}$ denotes the SYK$_4$ Hamiltonian of the left system. The effective temperature $\beta_{\text{eff}}$ scales as $J(\mu/J)^{2/3}$ at small $\mu/J$ [5]. There are $2^N$ Majorana strings $\Psi_{\boldsymbol{v}} = \Psi_{\boldsymbol{v}_0,L}\Psi_{\boldsymbol{v}_0,R}$ that stabilize the state at $\beta_{\text{eff}} = 0$, satisfying $\Psi_{\boldsymbol{v}}|\text{EPR}\rangle = |\text{EPR}\rangle$. Here, $\boldsymbol{v}_0$ is an $N$-dimensional vector. At finite $\beta_{\text{eff}}$, averaging over these operators gives

$$\overline{c_{\boldsymbol{v}}} = \overline{\langle\text{TFD}|\Psi_{\boldsymbol{v}_0,L}\Psi_{\boldsymbol{v}_0,R}|\text{TFD}\rangle} = \frac{(Z_{\beta_{\text{eff}}/2,\text{SYK}})^2}{2^{N/2}Z_{\beta_{\text{eff}},\text{SYK}}} = 2^{-\frac{N}{2}} e^{S_{2,\text{SYK}}(\beta_{\text{eff}}/2)}. \quad (16)$$

Neglecting (1) fluctuations in the differences among these operators and (2) contributions from other operators, this leads to the following expression

$$M_2 = -\ln\left(2^{-N}\sum_{\boldsymbol{v}} \overline{c_{\boldsymbol{v}}}^4\right) \approx 2N\ln 2 - 4S_{2,\text{SYK}}(\beta_{\text{eff}}/2). \quad (17)$$

This result approaches $M_2 = 0$ as $\mu \to \infty$, where the TFD state becomes a stabilizer state. On the contrary, we have $M_2 = N(2\ln 2 - 4s_0)$, with the zero-temperature entropy density of the SYK model given by $s_0 \approx 0.2324$ [6].

Next, we provide a direct proof of this result for the TFD state using the path-integral formalism, drawing on our understanding of the SRE for thermal density matrices of a single SYK model. The path-integral representation is shown in Fig. S1, where the solid lines represent the imaginary-time evolution governed by $H_{\text{SYK}}$. After reorganizing the fields, the path integral consists of four disconnected branches, each formed by joining two different replicas. For each mode $\psi_{L,j}$ or $\psi_{R,j}$, there is a auxiliary Ising spin $\sigma_{L,j}$ or $\sigma_{R,j}$, inserted with an imaginary separation of $\beta_{\text{eff}}/2$. The action can be derived by a series of observations:

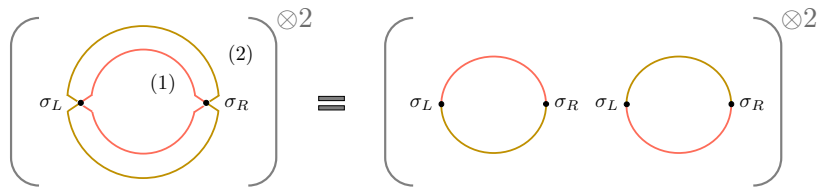

FIG. S1. Illustration of the path-integral representation of the stabilizer Rényi entropy for the TFD state. Here, we show only the contours for replicas (1) and (2), leaving those for (3) and (4) implicit.

1. Since the system describes the low-energy limit of the coupled SYK model, the path-integral exhibits the symmetry $\sigma_{L/R} \to -\sigma_{L/R}$ and

$$\big(\psi_{L/R}^{(\alpha_1)}, \psi_{L/R}^{(\alpha_2)}\big) \to -\big(\psi_{L/R}^{(\alpha_1)}, \psi_{L/R}^{(\alpha_2)}\big), \tag{18}$$

   for chosen $\alpha_1 \in \{1,2\}$ and $\alpha_2 \in \{3,4\}$. Assuming this symmetry is not spontaneously broken, we have $G_{ss'}^{(\alpha\beta)} = \delta^{\alpha\beta} G_{ss'}$ and $\Sigma_{ss'}^{(\alpha\beta)} = \delta^{\alpha\beta} \Sigma_{ss'}$, so that four different contours are completely decoupled. We can therefore focus on a single contour.

2. After introducing the bilocal fields, modes with different $j$ also decouple. For a given mode $j$ on a single path-integral contour, if $\sigma_L = \sigma_R$, the path integral has an anti-periodic boundary condition, as in the usual thermal partition function. However, when $\sigma_L = -\sigma_R$, the Majorana field experiences a periodic boundary condition. This is analogous to the calculation of the Witten index, which requires inserting the fermion parity operator $(-1)^F$. For a single mode, this can be implemented by inserting $\prod_\alpha \psi^{(\alpha)}(0)$ into a path integral with anti-periodic boundary conditions, similar to Eq. (13). Following a similar argument, this term does not contribute to the action and can be safely neglected. Therefore, we can impose $\sigma_L = \sigma_R$ for each mode $j$.

3. Finally, after imposeing $\sigma_L = \sigma_R$, each contour again takes exactly the same form as the norm of a KM pure state with inverse temperature $\beta_{\text{eff}}/2$. Therefore, we expect the action $S_{\text{SRE}} = 8 S_{\beta_{\text{eff}}/2, \text{SYK}} + \text{cons.}$. Combining with the normalization of Eq. (15) and determine the additive constant by $\beta_{\text{eff}} = 0$, we find $M_2 = 2N \ln 2 - 4 S_{2,\text{SYK}}(\beta_{\text{eff}}/2)$ as expected. Here, we use $S_{2,\text{SYK}}(\beta_{\text{eff}}/2) = S_{\beta_{\text{eff}},\text{SYK}} - 2 S_{\beta_{\text{eff}}/2,\text{SYK}}$.


\fi
  
\end{document}